\documentclass[aps,prb,twocolumn,showpacs,amsmath]{revtex4}
\usepackage{latexsym}
\usepackage{amssymb}
\usepackage{graphicx}
\usepackage{bm}% bold math
\usepackage[english]{babel}
\usepackage[latin1]{inputenc}
\usepackage{amsmath}
\usepackage{amsfonts}

\newcommand{\pdag}{{\phantom{\dagger}}}
\newcommand{\bq}{\begin{equation}}
\newcommand{\eq}{\end{equation}}
\newcommand{\bn}{\begin{eqnarray}}
\newcommand{\en}{\end{eqnarray}}

\begin{document}

\title{Exotic odd-even parity effects in transmission phase, (Andreev) conductance, and shot noise of a dimer atomic chain by %%@
topology}

\author{Bing Dong}
\affiliation{Key Laboratory of Artificial Structures and Quantum Control (Ministry of Education), Department of Physics and %%@
Astronomy, Shanghai Jiaotong University, 800 Dongchuan Road, Shanghai 200240, China}
\affiliation{Collaborative Innovation Center of Advanced Microstructures, Nanjing, China}

\author{X. L. Lei}
\affiliation{Key Laboratory of Artificial Structures and Quantum Control (Ministry of Education), Department of Physics and %%@
Astronomy, Shanghai Jiaotong University, 800 Dongchuan Road, Shanghai 200240, China}
\affiliation{Collaborative Innovation Center of Advanced Microstructures, Nanjing, China}

\begin{abstract}

We investigate the transport properties through a finite dimer chain connected to two normal leads or one normal and one %%@
superconductor (SC) leads. The dimer chain is described by the Su-Schrieffer-Hegger model and can be tuned into a topologically %%@
nontrivial phase with a pair of zero-energy edge states (ZEESs). We find that if the dimer chain is of nontrivial topology, (1) %%@
it will show apparent but opposite odd-even parity of the number of sites, in comparison with the topologically trivial and plain %%@
chains, in the (Andreev) transmission probability at the Fermi energy (i.e. the conductance and the Andreev conductance), the %%@
noise Fano factor in the zero bias limit, and even the transmission phase due to the coupled ZEESs; (2) the ZEES can determine %%@
appearance of the Andreev bound states at the site connected to the SC lead, and thereby induces a nonzero-bias-anomaly in the %%@
Andreev differential conductance of the hybrid junction; (3) the transmission phase of the normal junction has a unique $2\pi$ %%@
continuous phase variation at the zero-energy resonant peak that is also different from the usual phase shift in resonant point %%@
in usual systems. 

\end{abstract}

\date{\today}

\pacs{73.63.Nm, 72.15.Rn, 03.65.Vf, 05.60.Gg}

\maketitle

\section{Introduction}

Much intention has been paid, in recent years, to the physics of topological state in solid-state physics.\cite{Hasan} The %%@
experimental realization of the topology in one-dimensional (1D) systems,\cite{Atala,Xiao} by measuring the quantized Zak %%@
phase,\cite{Zak} has stimulated over again extensive investigation on the Su-Schrieffer-Hegger (SSH) model,\cite{Su} since this %%@
simple system has two topologically different phases and thus is a $Z_2$ topological %%@
insulator.\cite{Ryu,Delplace,Lang,Kraus,Li,Ganeshan,Gomez,Asboth} The two topological phases can be distinguished via the %%@
presence or absence, controlled by tuning two different dimerization strengths, of twofold degenerate zero-energy edge states %%@
(ZEESs) under the open boundary condition (OBC).\cite{Ryu,Delplace} It is therefore an intriguing issue to provide experimental %%@
confirmation for the existence of the ZEESs in the topological dimer chain. Recently, transport properties have been studied to %%@
sign the edge states of a finite dimer chain, when it is connected to two normal leads.\cite{Benito,Niklas,Ruocco} These works %%@
have discussed incoherent tunneling of the chain subject to a large bias voltage\cite{Benito} and/or a high frequency ac electric %%@
field.\cite{Niklas,Ruocco} In this paper, we examine the linear and nonlinear (Andreev) tunneling in the coherent regime for both %%@
a normal junction and a hybrid junction involving superconductor (SC).   

\section{Model of dimer chain}

We consider a 1D dimer lattice of $N$ sites, which contains two sublattices $a$ and $b$ in each unit cell with alternatingly %%@
modulated nearest-neighbor hopping amplitudes between them. It is just the celebrated SSH Hamiltonian,
\bq
H_{1D} = \sum_{j\sigma} \left ( t_1 d_{a\sigma,j}^{\dagger} d_{b\sigma,j}^\pdag + t_2 d_{a\sigma,j+1}^\dagger d_{b\sigma,j}^\pdag %%@
+ {\rm H.c.} \right ), \label{SSH}
\eq
where $d_{a\sigma,j}^\dagger$ $(d_{a\sigma,j})$ and $d_{b\sigma,j}^\dagger$ $(d_{b\sigma,j})$ are the fermion creation %%@
(annihilation) operators of electrons on the sublattices $a$ and $b$ of the $j$th unit cell with spin-$\sigma$, respectively. %%@
$t_{1(2)}=t_0\mp \delta t$ denote the hopping amplitudes in the unit cell and between two adjacent unit cells, $\delta t$ being %%@
the dimerization strength. Notice that the dimer chain showing topologically either trivial or nontrivial property can be easily %%@
controlled by simply tuning the relative strength of the intracell-to-intercell couplings, $\delta %%@
t$.\cite{Ryu,Delplace,Gomez,Asboth}
Throughout we will measure all energies in units of $t_0$ and use units $e=\hbar=k_B=1$.

\section{Normal junction}

In this paper, we propose two kinds of transport measurements for detecting the zero-mode by connecting the left and right ends %%@
of the dimer chain respectively to two electronic reservoirs. We consider the first setup that two leads are both normal metals %%@
(NCN junction), $H_{\eta} = \sum_{ {\bf k}\sigma} (\varepsilon_{\eta {\bf k}\sigma} -\mu_\eta) c_{\eta {\bf k}\sigma}^\dagger %%@
c_{\eta {\bf k}\sigma}^\pdag$, where
$c_{\eta{\bf k}\sigma}^\dagger$ ($c_{\eta{\bf k}\sigma}$) is the creation (annihilation) operator of an electron with momentum %%@
${\bf k}$ and spin-$\sigma$, energy $\varepsilon_{\eta {\bf k}}$, and chemical potential $\mu_\eta$ in the lead $\eta$ %%@
($\eta=\{L,R\}$). The tunneling Hamiltonian for the coupling between the chain and the leads are
\bq
H_{T} = \sum_{{\bf k}\sigma} ( \gamma_{L} c_{L {\bf k}\sigma}^\dagger d_{a\sigma,1} + \gamma_{R} c_{R {\bf k}\sigma}^\dagger %%@
d_{a(b)\sigma,N}+ {\rm H.c.} ).
\eq
The right end site of the chain could be sublattice either $a$ or $b$ depending on that the number $N$ of the sites is even or %%@
odd.
Here $\gamma_{\eta}$ describes the tunnel-coupling matrix element between the QD and lead $\eta$ and the corresponding coupling %%@
strength is defined as $\Gamma_\eta = 2\pi \sum_{k} |\gamma_\eta|^2 \delta(\omega-\varepsilon_{\eta {\bf k}})$. Without loss of %%@
generality, we use the wide band limit and assumed that $\Gamma_L=\Gamma_R=\Gamma$ is independent of energy to avoid undesirable %%@
effects from the conduction band edge. In addition, in order not to disturb the zero-energy mode of the dimer chain as far as %%@
possible, we set the tunnel-coupling $\Gamma=0.1$ in our following calculations to guarantee that it is much weaker than hopping %%@
amplitude, $\Gamma\ll t_0$.

Applying the nonequilibrium Green function (NGF) method, we can evaluate the current from the left lead to the chain and its shot %%@
noise as\cite{Haug}
\bn
I &=& \int \frac{d\omega}{\pi} T(\omega) \left ( f_R- f_{L} \right), \label{current} \\
S &=& 2\int \frac{d\omega}{\pi}\left \{ T(\omega) \left [ f_R (1- f_{L} ) + f_L (1-f_{R}) \right ] \right. \cr
&& \left. - T(\omega)^2 [ f_R - f_{L}]^2 \right \}. \label{sn}
\en
with the Fermi distribution $f_{\eta}=[e^{(\omega - \mu_\eta)/T}-1]^{-1}$ at the temperature $T$ and the transmission probability %%@
$T(\omega)={\Gamma_L \Gamma_R} |G_{1N}^r(\omega)|^2$.
Here, $G_{1N}^r(\omega)$ is the retarded GF between the first and last sites of the chain.

It is seen that the transport properties are completely determined by the transmission coefficient. Therefore we first examine %%@
how the transmission probability evolves when the atomic chain changes from linear to topological. In Fig.~1(a), we plot the %%@
transmission spectrum $T(\omega)$ as functions of the number of sites for the dimer chains with $\delta t=0.2$. It is observed %%@
that the transmission probability at $\omega=0$ shows (1) a nearly perfect transmission even for the even number of sites, %%@
$N=16$, but (2) a rapid decrease with increasing length of the chain, and (3) on the contrary, a nearly perfect reflection for %%@
the case of odd-site dimer chain. These behaviors are clearly different from those of the plain chain ($\delta t=0$). In fact, we %%@
can obtain the exact analytical expressions for $T(0)$.\cite{Zeng,Kim} For the plain chain, the transmission probability has the %%@
well-known odd-even parity dependence: $T(0)=16\Gamma_L \Gamma_R t_0^2/(\Gamma_L \Gamma_R+ 4t_0^2)^2\simeq 0$ for an even-site %%@
chain and $T(0)=4\Gamma_L \Gamma_R/(\Gamma_L+\Gamma_R)^2=1$ for an odd-site chain. In contrast, for the case of dimer chain, we %%@
have
\bq
T(0)=
\begin{cases}
\displaystyle\frac{(\Gamma/t_2)^2 (t_2/t_1)^N}{[(\Gamma/2t_2)^2(t_2/t_1)^N +1 ]^2}, & \text {$N$ is even}; \\
\displaystyle\frac{4(t_2/t_1)^{(N-1)}}{[(t_2/t_1)^{(N-1)} +1 ]^2}, & \text {$N$ is odd},
\end{cases}
\eq
showing an opposite odd-even parity. For instance, the dimer chain with $\delta t=0.2$ has a nearly perfect transmission at %%@
$\omega=0$ only if $N=16$, whereas it has a perfect reflection if $N$ is odd.

\begin{figure}[t]
\includegraphics[height=4.5cm,width=8cm]{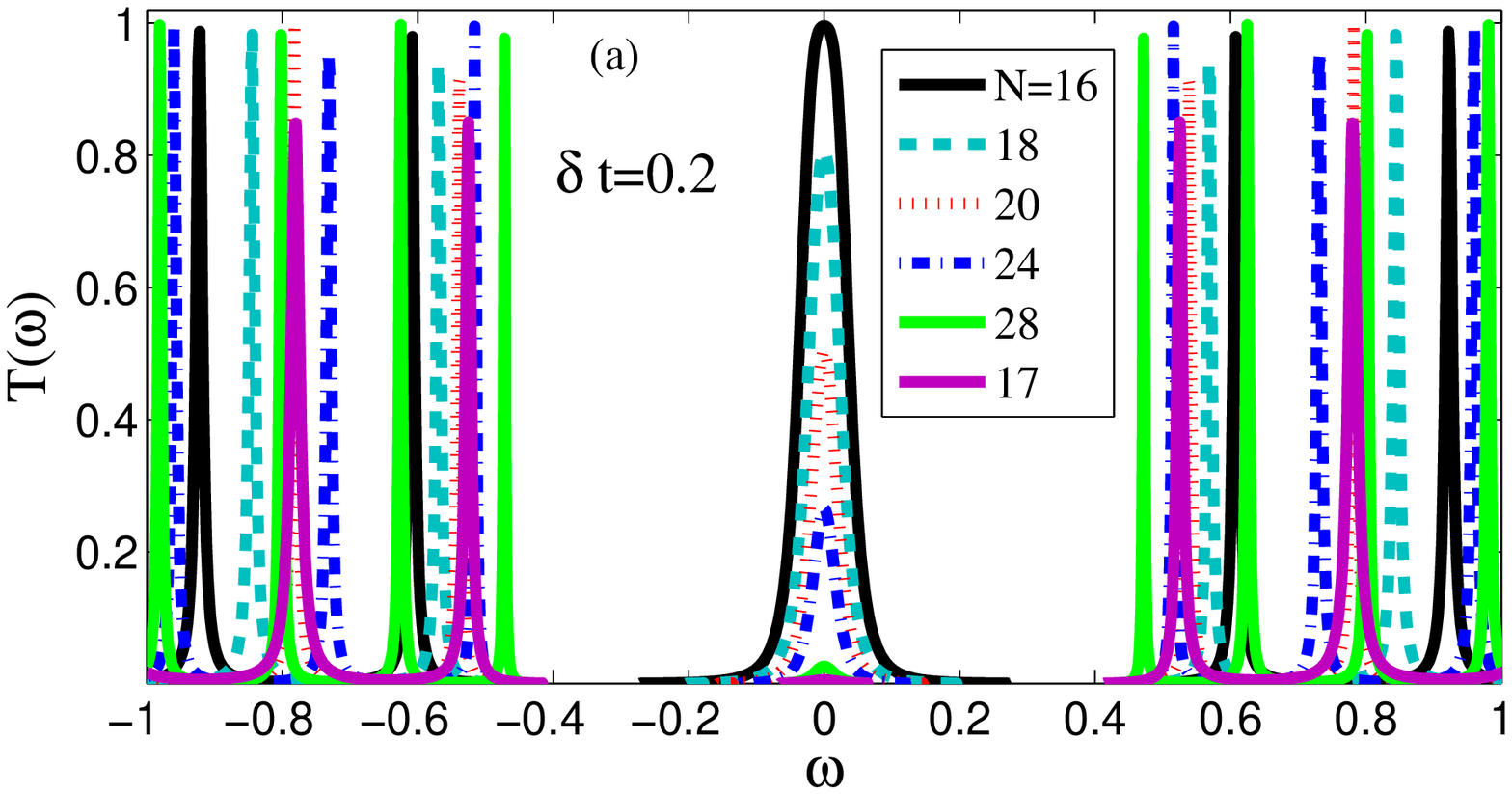}
\vspace{3mm}

\includegraphics[height=4cm,width=8cm]{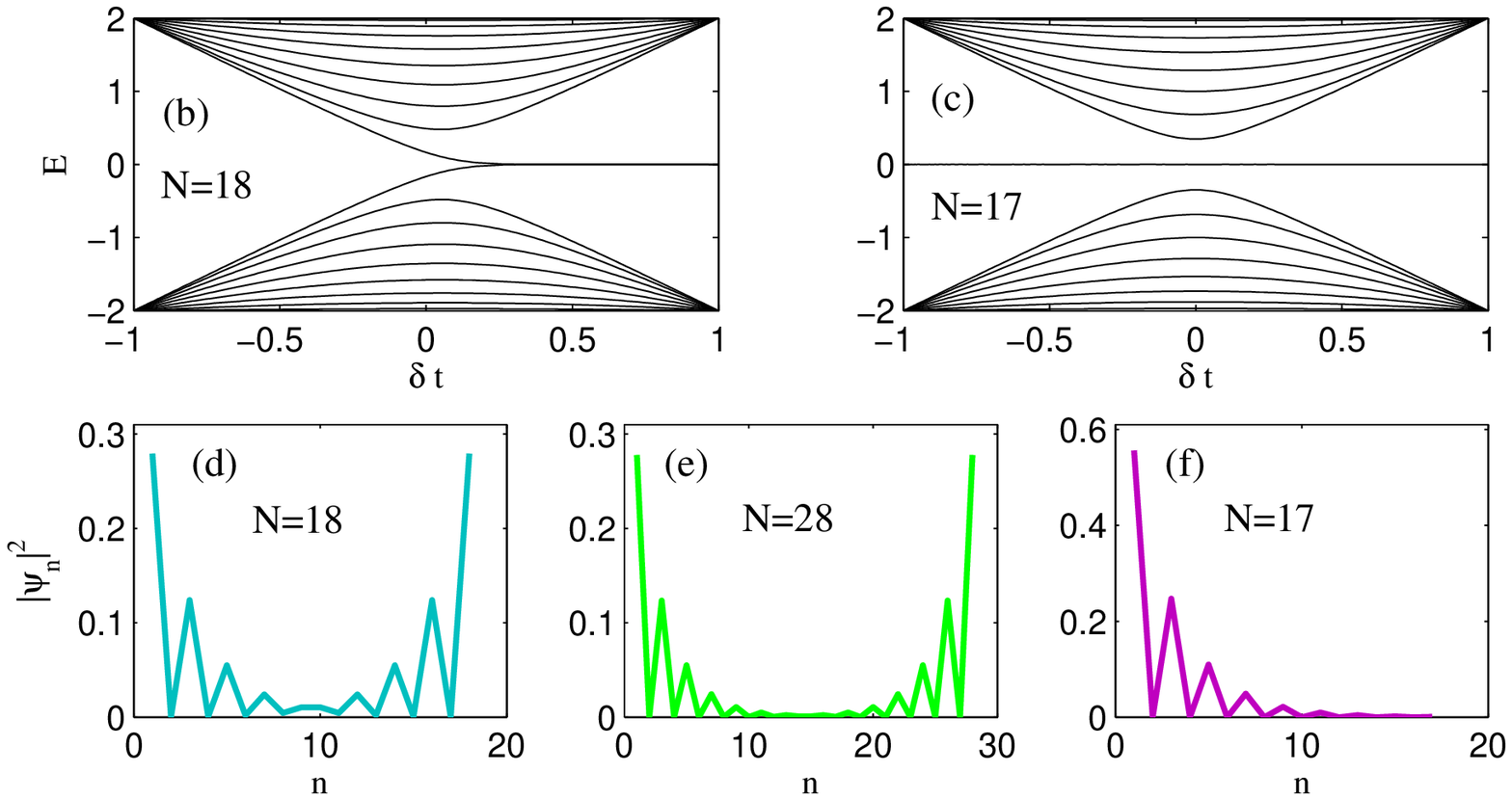}
\caption{(Colour online) (a) Dependence of the transmission probability on the number of atoms for the dimer chains with $\delta %%@
t=0.2$. (b,c) Energy spectrum of the dimer chains with even number $N=18$ and odd number $N=17$ of sites, respectively. (d,e,f) %%@
Norm of wave functions of the zero-energy modes at each site for the even-site short chain $N=18$, long chain $N=28$, and %%@
odd-site chain $N=17$.}
\label{fig1}
\end{figure}

These peculiar behaviors can be ascribed to the appearance of the ZEESs of the dimer chain. As shown in Fig.~1(b) for $N=18$, the %%@
dimer chain has a topologically nontrivial phase in the regime of $\delta t>0$ characterized by the presence of zero-energy %%@
states under the OBC, whereas no edge states exist in the regime of $\delta t<0$. From the distribution probability of the %%@
zero-energy states along the chain in the case of topological regime, $\delta t=0.2$ [Figs.~1(d,e)], we find that these states %%@
are indeed most probably occupied at the two endpoints of the chain. In addition, we find that the overlap integral of the wave %%@
function between the two coupled ZEESs can not be negligible for the short chain, $N=18$ [Fig.~1(d)], which induces a nonzero %%@
coupling between the two edge states. As a result, electron can be transferred coherently from one end to the other via the two %%@
ZEESs when the energy of the incident electron is zero, rather than via the scattering states as the odd-site plain chain does. %%@
Nevertheless, the overlap integral is quite sensitive to the chain length. It decays rapidly and becomes infinitesimal for the %%@
long chain, for instance, $N=28$ [Fig.~1(e)]. In this case, the two ZEESs become localized at the two endpoints respectively, and %%@
the transport channel closes. Moreover, for the odd-site chain, one ZEES exists at the whole regime of $\delta t$ [Fig.~1(c)]. %%@
But this state is always localized at one end of the chain [Fig.~1(f)] (the left end for $\delta t>0$ whereas the right end for %%@
$\delta t<0$), except for the plain chain ($\delta t=0$). Of course, the single localized edge state can not support electron %%@
tunneling, indicating strong localization in the odd-site dimer atomic wire.

\begin{figure}[htb]
\includegraphics[height=4.5cm,width=8.5cm]{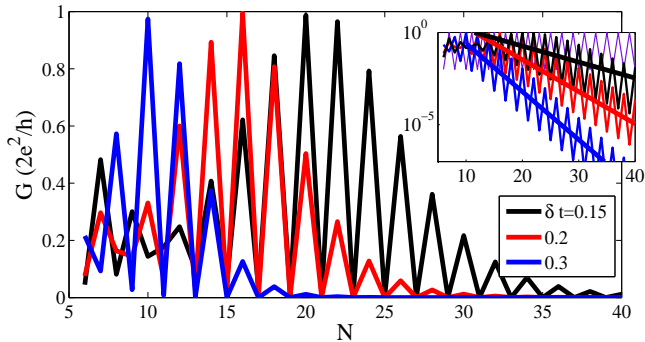}
\caption{(Colour online) The zero-temperature conductance as a function of the chain length for various $\delta t=0.15$ (black %%@
line), $0.2$ (red line), and $0.3$ (blue line). Inset: The conductance of the topological chain shows a perfect exponential decay %%@
with a decay coefficient proportional to $2\delta t/t_0$. While for the plain chain, no exponential decay is found (purple %%@
line).}
\label{fig2}
\end{figure}

The transmission probability at the Fermi energy can be detected in experiments by measuring the linear conductance,
$G=(2e^2/h) T(0)$, at zero temperature. We then calculate $G$ and plot them, in Fig.~2, as functions of chain length for various %%@
dimerization strengths. It is evident that the dimer chain displays the opposite odd-even parity to the plain chain. Moreover, %%@
the conductance of the dimer chain shows an exponential decay with increasing chain length, scaled properly as $\exp(-2N\delta %%@
t/t_0)$, while the plain chain does not. This fact confirms that the long dimer chain is an ideal Anderson insulator. In %%@
addition, the shot noise Eq.~(\ref{sn}) can be approximated as $S=(4e^2/h) T(0) [1-T(0)] eV$ in the limit of small bias voltage, %%@
$V=\mu_L-\mu_R$, at zero temperature. Therefore, it is expected that the Fano factor $F=S/2I$ of the dimer chain will also %%@
display the different odd-even behavior from the plain chain, as shown in Fig.~3.

\begin{figure}[htb]
\includegraphics[height=4.5cm,width=8.2cm]{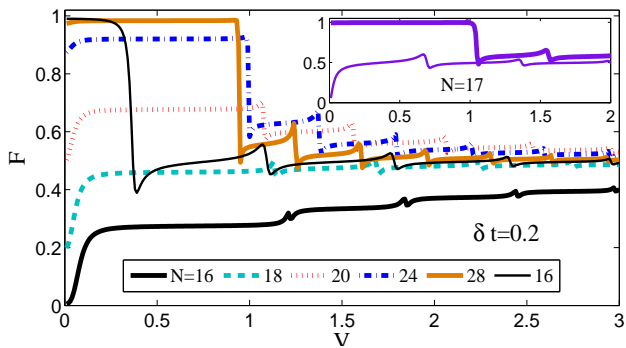}
\caption{(Colour online) The Fano factor $S/2I$ as a function of the bias voltage $V$ for different chain lengths with $\delta %%@
t=0.2$ (thick lines) and $\delta t=0$ (thin lines) at zero temperature.}
\label{fig3}
\end{figure}

\begin{figure}[t]
\includegraphics[height=4.5cm,width=8.5cm]{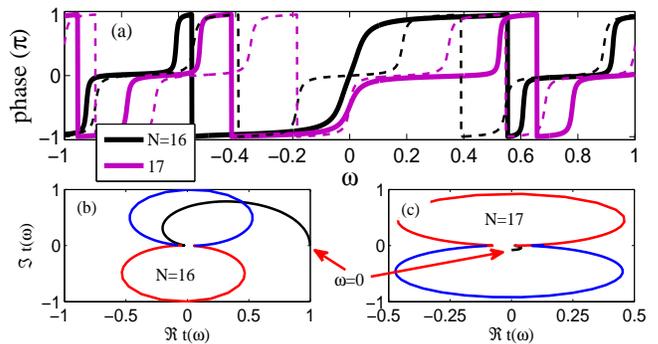}
\caption{(Colour online) (a) Transmission phase as functions of the electron energy for the even- and odd-site chains with %%@
$\delta t=0.2$ (thick solid lines) and $0$ (thin dashed lines). (b) Trajectory of the transmission amplitude for the $N=16$ dimer %%@
chain. The black line depicts the path of the electron energy range from $\omega=0$ to $0.55$, the red line for $\omega=0.55\sim %%@
0.86$, and the blue line for $\omega=0.86\sim 1.0$. (c) Trajectory of the transmission amplitude for the $N=17$ dimer chain for %%@
$\omega=0\sim 0.45$ (black line), $\omega=0.45 \sim 0.7$ (red line), and $\omega=0.7 \sim 1.0$ (blue line). The arrows indicate %%@
the starting point of the two paths.}
\label{fig4}
\end{figure}

We now analyze the transmission phase since it provides a complementary information to depict transmission coefficient, %%@
$t(\omega)=\sqrt{\Gamma_L \Gamma_R} G_{1N}^r(\omega)$.\cite{Lee,Taniguchi} For the plain chain, irrespective of the number $N$ of %%@
sites, the transmission resonant peaks are all out of phase, i.e. the phase increase continuously but rapidly by $\pi$ from one %%@
resonant peak to the next, as seen by the dashed thin lines in Fig.~4(a).\cite{Lee,Taniguchi,Zhai}
The odd-site dimer chain has the similar phase variation behavior for both the bulk states and the ZEES, even though the %%@
transmission amplitude is nearly zero at $\omega=0$. This behavior is illustrated in Fig.~4(c), showing that the path of %%@
$t(\omega)$ as a function of energy encircles the origin of the complex plane but always stay away from it. As the energy of %%@
incident electron is close to the bulk states, the circle radii are nearly equal to $1$ [red and blue lines in Fig.~4(c)], which %%@
are the typical trajectory of transmission amplitude for a single-channel symmetric double-barrier %%@
structure.\cite{Taniguchi,Zhai} Nevertheless, the radius of the trajectory becomes very small near the edge state [black line in %%@
Fig.~4(c)]. While for the case of even-site dimer chain, transmission amplitude starts from the positive real axis with $|t|=1$ %%@
and then evolves along the black line as seen in Fig.~4(b) to approach the origin. This indicates a continuous phase variation %%@
from $0$ to $2\pi$ when the energy of incident electron sweeps through the zero-energy resonance, and manifests that the two %%@
sides of the zero-energy resonance are in phase, which is just opposite to the regular resonances in the plain chain and even the %%@
resonances via the bulk states in the dimer chain. It can thereby be regarded as a unique signal for the appearance of the ZEESs.

\section{Hybrid junction}

We turn to the second transport setup at which the right lead is replaced with a SC (NCS junction), $H_R=\sum_{ {\bf k}\sigma} %%@
\varepsilon_{R {\bf k}} c_{R {\bf k}\sigma}^\dagger c_{R {\bf k}\sigma}^\pdag  + \sum_{{\bf k}} (\Delta c_{R {\bf %%@
k}\downarrow}^\dagger c_{R -{\bf k}\uparrow}^\dagger + {\rm H.c.})$, where $\Delta$ is the SC gap. Since we are interested only %%@
in the Andreev tunneling in the subgap region, $\Delta$ can be set as the biggest energy scale and thereby the role of the SC %%@
lead is solely to induce $s$-wave pairing for the site connected to the SC lead, i.e. the right end site in the NCS junction. To %%@
account for the proximity effect, we make use of the site$\otimes$Nambu space, $\psi=(d_{a\uparrow,1},\cdots, %%@
d_{a(b)\uparrow,N},d_{a\downarrow,1}^\dagger,\cdots,d_{a(b)\downarrow,N}^\dagger)^T$, to rewrite the effective Hamiltonian for %%@
the SSH model in matrix form as $\widetilde H_{1D}=\frac{1}{2} \psi^\dagger {\cal K} \psi$, where the $2N\times 2N$ matrix ${\cal %%@
K}$ is given by
\bq
{\cal K}=\left (
\begin{array}{cc}
K_{0} & K_{S} \\
K_{S} & -K_0
\end{array}
\right ),
\eq
in terms of the two $N\times N$ matrices,
\bq
K_0=\left (
\begin{array}{cccccc}
0 & t_1 & 0 & \cdots & \cdots & 0 \\
t_1 & 0 & t_2 & 0 & \cdots & 0 \\
0 & t_2 & 0 & \ddots & \ddots & \vdots \\
\vdots & \ddots & \ddots & \ddots & \ddots & \vdots \\
\vdots & \ddots & \ddots & \ddots & \ddots & t_{1(2)} \\
0 & \cdots & \cdots & \cdots & t_{1(2)} & 0
\end{array}
\right ),
\eq
and $K_S$ having only one nonzero element for the right end site of the chain, $K_{S,NN}=\Gamma_S/2$ (Here we use $\Gamma_S$ %%@
instead of $\Gamma_R$ to signify the SC hybrid effect). Without loss of generality, we set $\Gamma_S=3\Gamma$ in the calculations %%@
such that a pair of the Andreev bound states (ABSs) is established unambiguously.
The retarded GF ${\cal G}^r(\omega)$ of the chain, defined as ${\cal G}^r(t,t')=-i\theta(t-t')\langle [\psi(t), \psi^\dagger %%@
(t')] \rangle$, can be formally given by the Dyson equation
\bq
{\cal G}^r(\omega)=\left [ \omega {\bm I}- {\cal K} - \left (
\begin{array}{cc}
\Sigma^r & 0 \\
0 & \Sigma^r
\end{array}
\right ) \right ]^{-1},
\eq
where ${\bm I}$ is a $2N\times 2N$ unit matrix and the $N\times N$ retarded self-energy $\Sigma^r$ is due to the coupling to the %%@
left norm lead. It has only one nonzero diagonal element, $\Sigma_{11}^r=-i\Gamma_L/2$.

\begin{figure}[b]
\includegraphics[height=3.6cm,width=8cm]{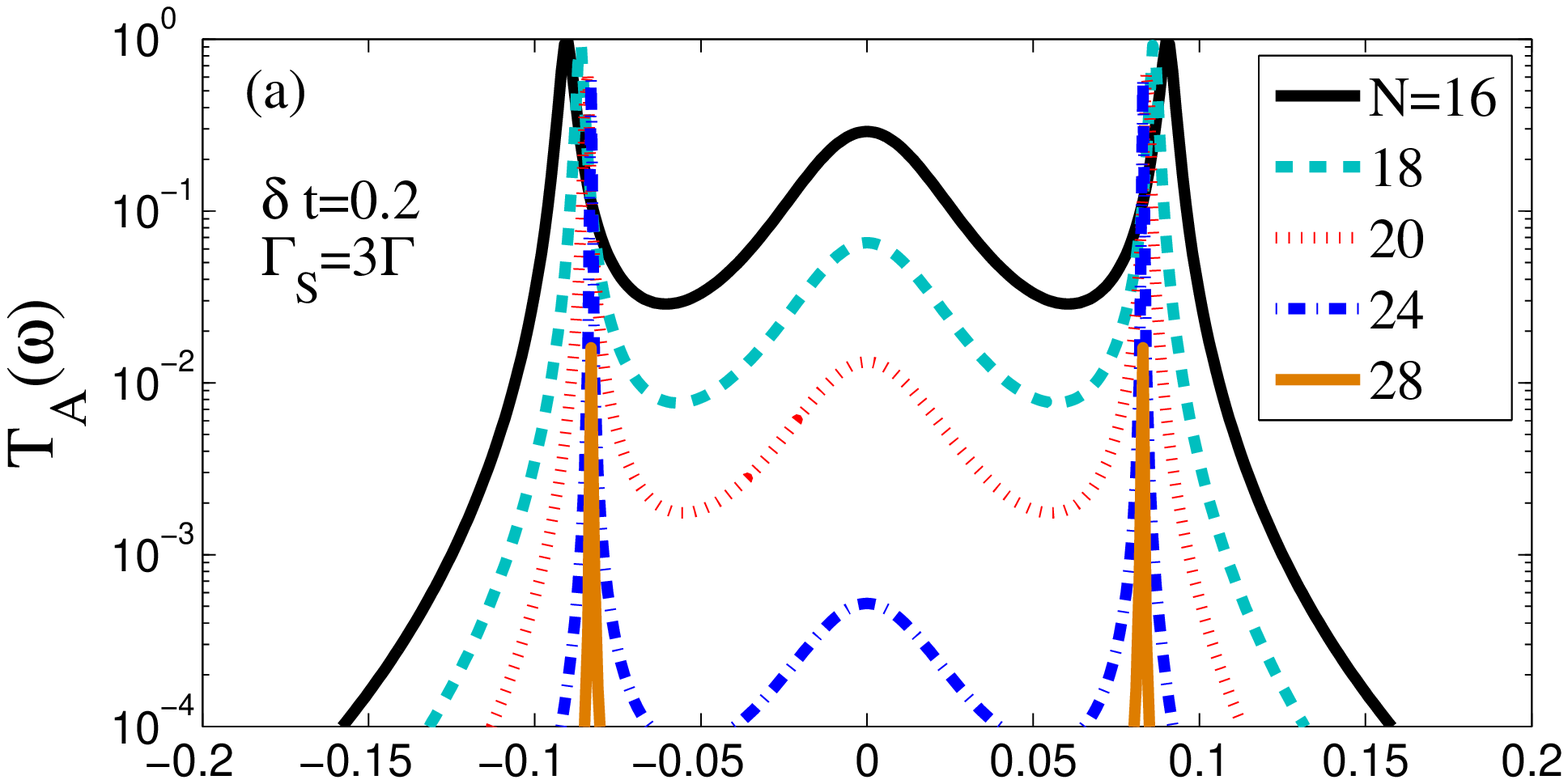}
\vspace{3mm}

\includegraphics[height=3.8cm,width=8cm]{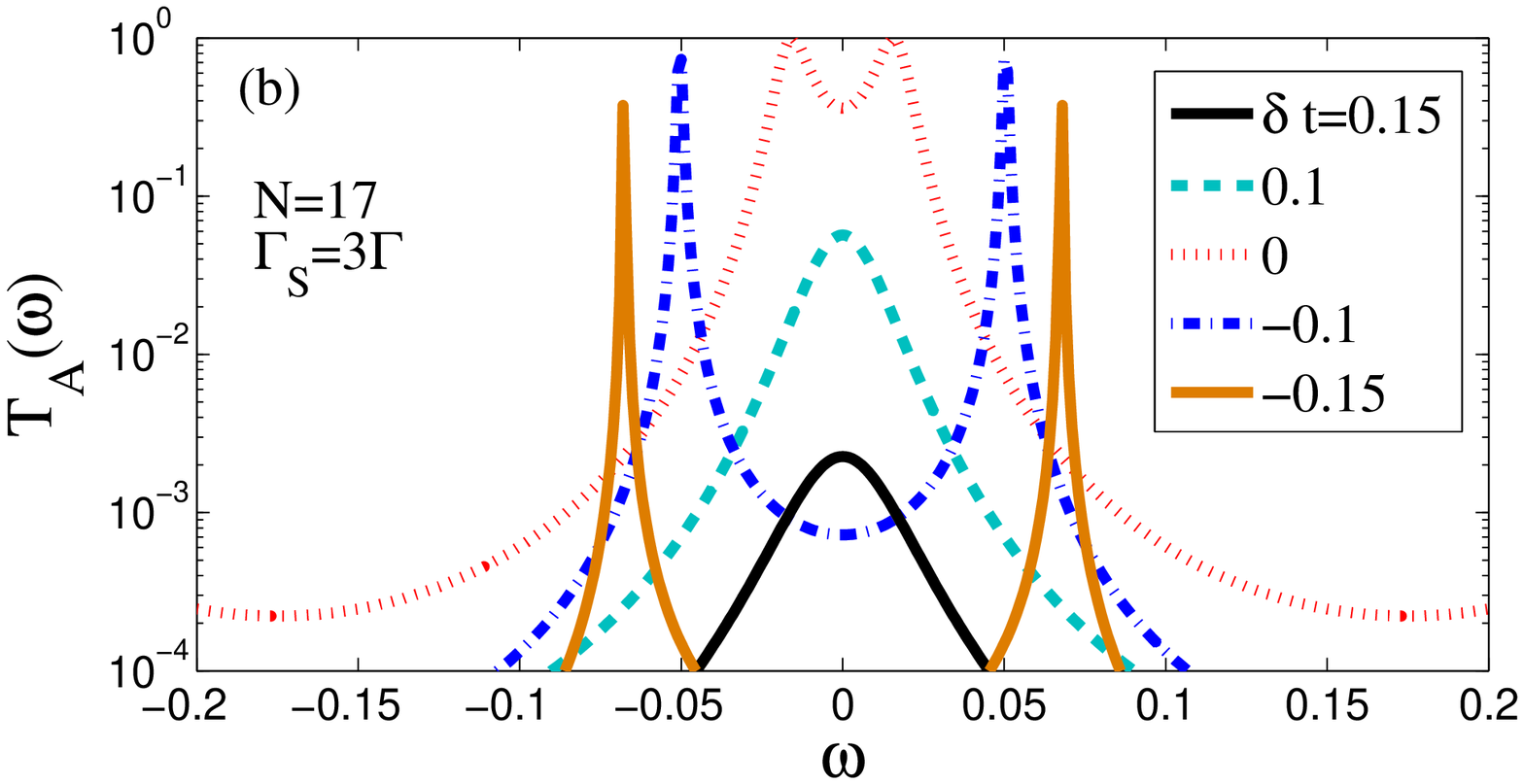}
\vspace{3mm}

\includegraphics[height=2.4cm,width=8cm]{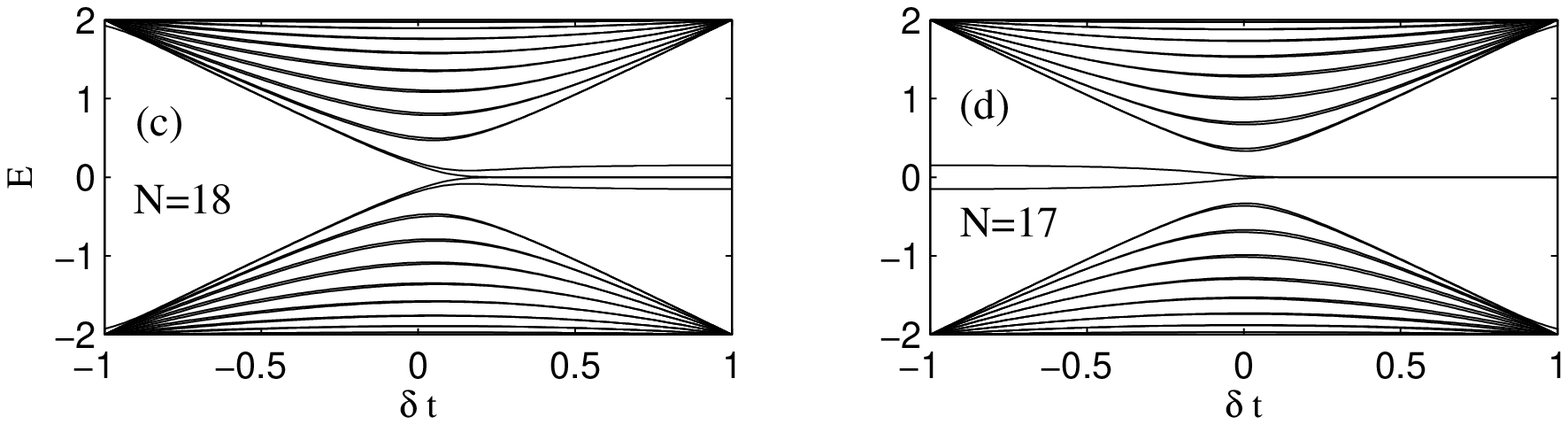}
\caption{(Colour online) (a) Dependence of the Andreev reflection spectrum on the even number of sites for the dimer chains with %%@
$\delta t=0.2$ and $\Gamma_S=3\Gamma$. (b) The Andreev reflection spectrum of the odd site $N=17$ with various dimerization %%@
strengths. (c,d) Energy spectrum of the dimer chains with the even site $N=18$ and the odd site $N=17$, respectively.}
\label{fig5}
\end{figure}

Employing NGF method, we can derive the Andreev current for the subgap tunneling as\cite{Kim,Dong}
\bq
I_A = - \int \frac{d\omega}{\pi} T_A(\omega) \left ( 1-f_L- f_{-L} \right),
\eq
with the hole distribution function $f_{-L}=[e^{(-\omega - \mu_L)/T}-1]^{-1}$, and the Andreev reflection probability
$T_A(\omega)=\Gamma_L^2 |{\cal G}_{1,N+1}^r(\omega)|^2$. In this kind of hybrid transport device, the chemical potential of the %%@
SC lead is usually set as a reference, and the external bias voltage is applied to the norm lead, $\mu_L=V$. The noise of the %%@
Andreev current can therefore be calculated at zero temperature as\cite{Dong,Muzykantskii}
\bq
S_A = 2 \int_{-V}^V \frac{d\omega}{\pi} \, T_A(\omega) [ 1- T_A(\omega) ]. \label{snA}
\eq

We then analyze the topological effect on the Andreev reflection spectrum $T_A(\omega)$ of the dimer chain.
We have the exact expressions for $T_A(0)$.\cite{Kim} The odd-even parity is also evident in the Andreev conductance of the plain %%@
chain as in the NCN junction: $T_A(0)=(\Gamma_L \Gamma_S t_0^2/2)^2 /[(\Gamma_L \Gamma_S)^2+ t_0^4]^2\simeq 0$ if $N$ is even, %%@
while $T(0)=4\Gamma_L^2 \Gamma_S^2/(\Gamma_L^2+\Gamma_S^2)^2=0.36$ if $N$ is odd. For the dimer chain, we obtain\cite{Kim}
\bq
T_A(0)=
\begin{cases}
\displaystyle\frac{(\Gamma_L\Gamma_S /2t_2^2)^2 (t_2/t_1)^{4N}}{[(\Gamma_L \Gamma_S/4t_2^2)^2(t_2/t_1)^{4N} +1 ]^2}, & \text{$N$ %%@
is even}; \\
\displaystyle\frac{\Gamma_L^2 \Gamma_S^2 (t_2/t_1)^{4N}}{4[(\Gamma_L/2)^2 (t_2/t_1)^{4N} + (\Gamma_S/2)^2 ]^2}, & \text{$N$ is %%@
odd}.
\end{cases}
\eq
An opposite odd-even parity is also found, for instance, $T_A(0)=0.29$ for the short even-site chain ($N=16$), while %%@
$T_A(0)\simeq 0$ for the $N=17$ chain with $\delta t=0.2$. More interestingly, we find from Fig.~5(a) that $T_A(\omega)$ 
exhibits triple peaks, say one zero-energy peak and two sharp peaks located equally at the left and right sides of the central %%@
peak, 
for the even-site dimer chain in the topologically nontrivial phase. In addition, the height of the two side peaks is nearly %%@
unity for short chains and decreases with increase of the chain length also but relatively much slowly compared with the central %%@
peak. The side peaks can be ascribed to the emergence of a pair of ABSs adhered to the ZEES due to SC proximity effect, as %%@
illustrated in Fig.~5(c). While in the topologically trivial case ($\delta t<0$), no ZEES means no ABS, and thus no side peaks. 
Another intriguing result is that for the odd-site dimer chain, $T_A(\omega)$ shows either a single central peak if $\delta t>0$ %%@
or two side peaks if otherwise [Fig.~5(b)]. This is because the isolated ZEES is located at the left end of the odd-site chain in %%@
the case of $\delta t>0$, thereby no electronic state exists at the right end to support the ABS; on the contrary, the degenerate %%@
single ZEES at the right end is splitting owing to the SC proximity effect and becomes a pair of ABSs in the case of $\delta %%@
t<0$[Fig.~5(d)].  

\begin{figure}[t]
\includegraphics[height=5cm,width=8.2cm]{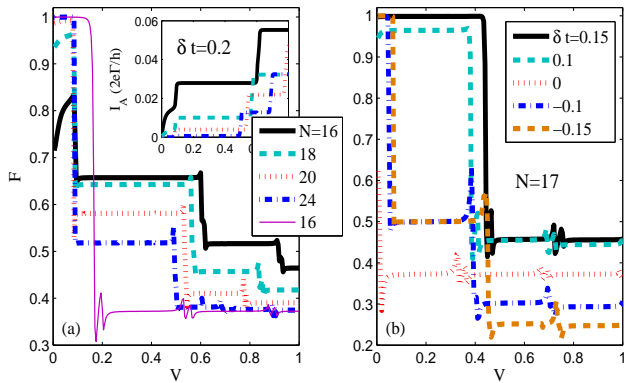}
\caption{(Colour online) The Fano factor of the Andreev current, $S_A/2I_A$ at zero temperature, as a function of the bias %%@
voltage $V$ for (a) different chain lengths with $\delta t=0.2$ at the topologically nontrivial case (the thin line denotes the %%@
result of $\delta t=0$), and (b) various $\delta t$ for the odd-site chain ($N=17$). Inset in (a) plots the corresponding %%@
$I_A$-$V$ curves.}
\label{fig6}
\end{figure}

The ZEES supported ABS can be detected through two quantities in the nonlinear regime. The first quantity, for example, is the %%@
nonzero-bias-anomaly in the differential Andreev conductance. Indeed, from the inset figure in Fig.~6(a), it is observed a upward %%@
jump in the Andreev current $I_A$-$V$ curves stemming from the sharp peak of the ABS for the short even-site dimer chains in the %%@
topologically nontrivial phase. Correspondingly, the ABS induces a sharp downward jump in the current noise, i.e. the Fano factor %%@
$F=S_A/2I_A$. Likewise, a rapid downward jump is also found for the noise of the odd-site chain, $N=17$, in the case of $\delta %%@
t<0$. While in the case of $\delta t>0$, the noise remains Poissonian, i.e. $F=1$, until the bulk state begins to contribute %%@
[Fig.~6(b)].    

\section{Summary}

Using the SSH tight-binding model, we have analyzed the topological effect on the transport properties of a 1D dimer chain when %%@
the chain is connected to two normal leads or to one normal and one SC leads by means of NGF method. It has been found that the %%@
topologically nontrivial chain possesses an opposite odd-even parity dependences of the (Andreev) conductance and the noise Fano %%@
factor in the zero-bias limit with respect to the number of sites, compared with results of the plain chain. Moreover, we have %%@
predicted a nonzero-bias-anomaly in the Andreev differential conductance of the NCS junction, and ascribed it to the emergence of %%@
ABSs due to the joint effects of the ZEES and the SC proximity effect. Besides, we have also analyzed the transmission phase of %%@
the NCN junction and found a unique $2\pi$ continuous phase variation at the zero-energy resonant peak only when the dimer chain %%@
is in the topologically nontrivial phase. Finally we would like to mention that we propose, in this paper, two simple transport %%@
experiments to detect the ZEES, which could provide useful information to identify the topological quantum phase transition in %%@
the 1D atomic wire.

\begin{acknowledgments}

This work was supported by Projects of the National Basic Research Program of China (973 Program) under Grant No. 2011CB925603, %%@
and the National Science Foundation of China under Grant No. 11674223.

\end{acknowledgments}

\end{document}